\documentclass[12pt,preprint]{aastex}
\begin{document}

\title{THE DISTRIBUTION, EXCITATION AND FORMATION OF COMETARY MOLECULES: METHANOL, METHYL CYANIDE AND ETHYLENE GLYCOL}

\author{Anthony J. Remijan\altaffilmark{1}, Stefanie N. Milam\altaffilmark{2,3},  
Maria Womack\altaffilmark{4}, A. J. Apponi\altaffilmark{2}, L. M. Ziurys\altaffilmark{2},  
Susan Wyckoff\altaffilmark{5}, M. F. A'Hearn\altaffilmark{6}, Imke de Pater\altaffilmark{7}, J. R. Forster\altaffilmark{7}, 
D. N. Friedel\altaffilmark{8}, Patrick Palmer\altaffilmark{9},
 L. E. Snyder\altaffilmark{8}, J. M. Veal\altaffilmark{8,10}, L. M. Woodney\altaffilmark{11}, 
\& M. C. H. Wright\altaffilmark{7}}

\altaffiltext{1}{National Radio Astronomy Observatory, 520 Edgemont Road, Charlottesville, VA 22901\\
email: aremijan@nrao.edu.}

\altaffiltext{2}{NASA Astrobiology Institute, Department of Chemistry, Department of Astronomy, and
Steward Observatory, 933 North Cherry Avenue, University of Arizona, Tucson, AZ 85721; email: 
lziurys@as.arizona.edu, aapponi@as.arizona.edu.}

\altaffiltext{3}{Current Address: NASA Ames Research Center, Astrophysics Branch MS 245-6, Moffett Field, 
CA 94035-1000; email: stefanie.n.milam@nasa.gov}

\altaffiltext{4}{Department of Physics and Astronomy, St. Cloud State University, St. Cloud, MN 56301;
mwomack@stcloudstate.edu.}

\altaffiltext{5}{Department of Physics and Astronomy, Arizona State University, Tempe AZ 85287;
wyckoff@asu.edu.}

\altaffiltext{6}{Department of Astronomy, University of Maryland, College Park MD 20742-2421
email: ma@astro.umd.edu}

\altaffiltext{7}{Department of Astronomy, University of California, Berkeley, CA 94720
email: imke@floris.berkeley.edu, rforster@astro.berkeley.edu, wright@astro.berkeley.edu}

\altaffiltext{8}{Department of Astronomy, 1002 W. Green St., University of Illinois, Urbana IL 61801
email: friedel@astro.uiuc.edu, snyder@astro.uiuc.edu}

\altaffiltext{9}{Department of Astronomy and Astrophysics, University of Chicago, Chicago, IL 60637
email: ppalmer@oskar.uchicago.edu}

\altaffiltext{10}{Current address: Southwestern College, 900 Otay Lakes Road, Chula Vista,
CA 91910; jveal@swccd.edu}

\altaffiltext{11}{Department of Physics, California State University, San Bernardino,
CA 92407; woodney@csusb.edu.}

\clearpage

\begin{abstract}

We present an interferometric and single dish study of small organic species toward
Comets C/1995 O1 (Hale-Bopp) and C/2002 T7 (LINEAR) using the BIMA interferometer at 
3 mm and the ARO 12m telescope at 2 mm.  For Comet Hale-Bopp, both the single-dish 
and interferometer observations of CH$_3$OH indicate an excitation temperature
of 105$\pm$5 K and an average production rate ratio $Q(CH_3OH)/Q(H_2O)$$\sim$1.3\% at 
$\sim$1 AU.  Additionally, the aperture synthesis observations of CH$_3$OH suggest
a distribution well described by a spherical outflow and no evidence of
significant extended emission. Single-dish observations of CH$_3$CN in
Comet Hale-Bopp indicate an excitation temperature of 200$\pm$10 K and a production rate ratio 
of $Q(CH_3CN)/Q(H_2O)$$\sim$0.017\% at $\sim$1 AU. 
The non-detection of a previously claimed transition of cometary (CH$_2$OH)$_2$ toward Comet Hale-Bopp 
with the 12m telescope indicates a compact distribution of emission, D$<$9$''$ ($<$8500 km).  
For the single-dish observations of 
Comet T7 LINEAR, we find an excitation temperature of CH$_3$OH of 35$\pm$5 K and a 
CH$_3$OH production rate ratio of $Q(CH_3OH)/Q(H_2O)$$\sim$1.5\% at $\sim$0.3 AU.
Our data support current chemical models that CH$_3$OH, CH$_3$CN and (CH$_2$OH)$_2$ 
are parent nuclear species distributed into the coma via direct
sublimation off cometary ices from the nucleus with no evidence of
significant production in the outer coma.  

\end{abstract}


\section{INTRODUCTION}

Over the last several years, complementary observations of comets with both single-dish telescopes and 
interferometric arrays, have shed new light on the abundance, production rate, distribution 
and formation of molecules in cometary comae.  Recently, Milam et al.\ (2006) employed this method to 
investigate the origins of cometary formaldehyde (H$_2$CO) by comparing observations with both the Arizona
Radio Observatory (ARO) 12m\footnote{The Kitt Peak 12m telescope is currently operated by Arizona
Radio Observatory (ARO), Steward Observatory, University of Arizona.}
single-dish telescope and the Berkeley-Illinois-Maryland Association (BIMA) array\footnote{Operated by the University 
of California, Berkeley, the University of Illinois, and the University of Maryland with support 
from the National Science Foundation.}. From this investigation 
toward Comet C/1995 O1 (Hale-Bopp), the authors clearly showed that the distribution and abundance of H$_2$CO 
must be coming from a source other than the comet nucleus. The extended source of H$_2$CO may be due to an 
enhanced production of this molecule, ($\sim$4$\times$ larger) from grains consisting of silicates and organic material 
(Milam et al.\ 2006), or possibly from the photodissociation of a larger parent species such 
as methanol (CH$_3$OH) (Hudson, Moore \& Cook 2005).  These complementary observations, however, 
are very difficult to obtain, as most of the molecular observations
toward comets are performed using solely single-dish telescopes or interferometric arrays, depending 
on the telescope available at the time as the comet approaches the inner solar system and perigee.

Multiple facility collaborations for cometary observations are necessary if we are to further 
understand the formation, excitation and distribution of organic cometary material.  Recent
cometary models of the abundance of organic material suggest Oort Cloud comets may have
had an origin in the giant planet region, where they were thermally and chemically processed, before being
ejected into the Oort Cloud (Weissman 1999).  Furthermore, observations comparing long- and short-period
comets suggest distinct classes - organic ``rich'', ``normal'', and ``depleted'' (Bonev et al.\ 2008;
Mumma et al.\ 2003).  However, the current sample size is too limited to accurately characterize each class of 
comets.  It should also be noted that many of the comets described as organic ``rich'' have undergone 
breakup or fragmentation events, thus exposing more internal, unprocessed material than an 
object just entering the inner solar system (Bonev et al.\ 2008). So,
the question persists, how does the classification of organic
``rich'', ``normal'' or ``depleted'' hold with regards to comets that
are for example, dust rich versus dust poor, having undergone 
a recent breakup, or bear high abundances of hypervolatile species? 

The organic species examined in this study, methanol (CH$_3$OH), methyl cyanide (CH$_3$CN) and 
ethylene glycol ([CH$_2$OH]$_2$) are believed to be parent species directly sublimated off the cometary 
ices in the nucleus.  Since the first reported single-dish detections of cometary CH$_3$OH towards Comet 
Brorsen-Metcalf (1989~o) and Comet Austin (1989~c1) (Colom et al.\ 1990), CH$_3$OH remains the most abundant 
large ($>$5 atoms) organic (containing a C-atom) molecule measured in cometary ices. In a recent review by Charnley \& Rodgers 
(2008), CH$_3$OH has a measured abundance range relative to H$_2$O between 1-4\%.  However, aperture 
synthesis observations of cometary CH$_3$OH are very rare and no high spatial resolution data of this molecule currently exists.
Similarly, single-dish observations of CH$_3$CN and (CH$_2$OH)$_2$ also suggest a purely nuclear 
origin (Biver et al.\ 2002, Crovisier et al.\ 2004). CH$_3$CN, as well as HC$_3$N, may be parent species that would 
increase the abundance of the CN radical as well as C$_2$ in the coma. Yet, current observations do not 
preclude the possibility of CH$_3$CN having its own parent species (Bockel{\'e}e-Morvan et al.\ 1987).  In addition, the 
high abundance of (CH$_2$OH)$_2$ observed in Comet Hale-Bopp at high spatial resolution also strongly
suggests a nuclear origin.  To date, complementary observations of the same transitions of either CH$_3$CN 
or (CH$_2$OH)$_2$ with another facility were not available to follow-up on the claimed distribution of 
these species in the coma and their possible formation pathways.

In this paper, we report on complementary observations of cometary CH$_3$OH, CH$_3$CN  and (CH$_2$OH)$_2$ 
taken with the BIMA array near Hat Creek, CA and the ARO 12m telescope on Kitt Peak, AZ toward Comets 
C/1995 O1 (Hale-Bopp) and C/2002 T7 (LINEAR).  From these observations, we continued to investigate 
the abundance, production rate, distribution and formation of molecules in comets by directly comparing 
the measurements of these organic species with different telescopes.  Our data supports the
hypothesis that CH$_3$OH, CH$_3$CN and (CH$_2$OH)$_2$ are primarily parent species, generated by 
direct sublimation of cometary ices contained in the nucleus.  

\section{OBSERVATIONS}

\subsection{BIMA Array}

Our observations of Comet Hale-Bopp with the BIMA array constitute one of the earliest detections 
of cometary CH$_3$OH with an interferometer.  These measurements were taken using the ``soft'' C configuration
of the BIMA array from 1997 March 24 to April 03. Data were acquired in the interferometric 
(cross-correlation) mode with nine antennas. The minimum baseline for these observations
was $\sim$15 m and the maximum baseline was $\sim$139 m. The average resulting full width
half power synthesized beam size was $\sim$8$''$$\times$10$''$ for the observations.  
The quasar 0102+584 was used to calibrate the antenna based gains and 
the absolute amplitude calibration of this source was based on planetary observations
and is accurate to within $\sim$20\%.  The spectral windows containing the transitions had
a bandwidth of 25 MHz and were divided into 256 channels for a spectral resolution
of 0.1 MHz per channel.  However, to increase the signal-to-noise ratio in each window,
the data were averaged over two channels giving an effective spectral resolution of
0.2 MHz per channel.  All data were first corrected to JPL ephemeris reference orbit
139.  The data were then combined and imaged using the MIRIAD software package (Sault,
Teuben \& Wright 1995).  To include all the data from the comet with multiple tracks,
the data were inverted in {\it u-v} space.  The resulting synthesized beamsize was determined
from the combined dataset.  The uvdata from multiple tracks were
combined to make images. Table 1 summarizes the observational parameters of the BIMA array observations. 

\subsection{12m Single Dish}

Observations at 2mm of CH$_3$OH, CH$_3$CN and (CH$_2$OH)$_2$ toward Comet Hale-Bopp were taken during three
observing runs on 1997 Mar 10, 11, and 20 using the ARO (at the time, NRAO\footnote{The National Radio 
Astronomy Observatory is a facility of the National Science Foundation operated under cooperative agreement by the Associated
Universities, Inc.}) 12m telescope.  Observations of 
CH$_3$OH toward Comet LINEAR were conducted on 2004 May 20 also using
the ARO 12m telescope.  Dual-channel SIS mixers were used for the 2 mm
observations, operated in single-sideband mode with $\sim$20 dB image rejection.  The backends employed for 
the presented observations were individual filter banks with 500 kHz resolution and the millimeter 
autocorrelator (MAC) with a resolution of 781 kHz. 
The spectral temperature scale was determined by the chopper-wheel method, corrected for
forward spillover losses, given in terms of $T_R^*$ (K).  The radiation temperature, $T_R$,
is then derived from the corrected beam efficiency, $\eta_c$, where  $T_R = T_R^*/\eta_c$.
A two body ephemeris program was used to determine the comet's position using the orbital
elements provided by D. Yeomans (1996, private communication) of JPL for Comet Hale-Bopp
and JPL ephemeris reference orbit 69 was used for the Comet LINEAR observations.  Focus and positional
accuracy were checked periodically on nearby planets or maser sources.  Data were taken
in the position-switched mode with the off position 30$'$ west in azimuth. Observing 
frequencies, dates, comet geocentric and heliocentric distances, beam size, the diameter of 
the projected beam size on the comet, and the corrected main beam efficiency ($\eta_c$) 
at the times of observations are listed in Table 2.

Table 3 lists the transitions and molecular line parameters of CH$_3$OH, CH$_3$CN and (CH$_2$OH)$_2$.
All spectral line data were taken from the Spectral Line Atlas of Interstellar Molecules (SLAIM) 
available at www.splatalogue.net (Remijan et al.\ 2008) and the Cologne Database for Molecular 
Spectroscopy (M{\"u}ller et al.\ 2005).

\section{RESULTS}

\subsection{BIMA Array}

Figure 1 shows the images and spectra of CH$_3$OH taken toward Comet Hale-Bopp with the BIMA array.  Figure 
1(a) shows the image of the J(K$_a$,K$_c$)=2(0,2)-1(0,1) E transition of CH$_3$OH averaged over 2 days,
1997 March 27 and 31.  Notice that the CH$_3$OH emission is slightly larger than the synthesized beam of 
the BIMA array (located at the bottom left of each image) and follows the general position angle 
of the synthesized beam.  The projected direction to the sun is shown by the line segment.
The coordinates are given in 
offset arcseconds centered on the comet nucleus. Figure 1(b) shows the spectra of 4 
CH$_3$OH transitions near 96.7 GHz (Table 3) including the J(K$_a$,K$_c$)=2(0,2)-1(0,1) E transition.  
The spectrum has been hanning smoothed over 3 channels.  The rms noise level is $\sim$0.1 Jy 
beam$^{-1}$ (indicated by the vertical bar at the left of the spectrum).  The spectral 
line labels correspond to the rest frequency located at the top left of the 
panel for a cometocentric velocity of 0 km~s$^{-1}$, denoted by a dashed line.
 
Figure 1(c) shows the image of the J(K$_a$,K$_c$)=3(1,3)-4(0,4) A$^+$
transition of CH$_3$OH averaged over 3 days, 1997 March 26 and April 2-3.  Figure 1(d) shows the spectrum of 
this CH$_3$OH transition near 107.0 GHz (Table 3).   Note that the two CH$_3$OH emission contours shown 
in figures (a) and (c) are centered on the predicted location of the comet nucleus with no apparent offset. 
The images are similar to each other and similar to the average of the March HCN BIMA array contours given 
by Veal et al.\ (2000) and the CO BIMA array contours given by Milam et al.\ (2006).  In addition, there is no
pronounced evidence for the jets or significant extended enhancement of CH$_3$OH beyond the 3$\sigma$ 
detection level ($\S$4.2). Furthermore, from figures 1(b) and (d), there is no apparent second velocity 
component of CH$_3$OH nor is the emission offset from the cometocentric rest velocity of 0 km s$^{-1}$.  

\subsection{12m Single Dish}

Figure 2 shows the spectra of the five $\lambda$ $\simeq$ 2mm transitions of CH$_3$OH detected 
toward Comet Hale-Bopp with the 12m telescope (Table 3).  The UT date and spectral line 
quantum numbers of each observed transition are located at the top of each panel.  
As in the BIMA spectra, the spectral line labels correspond to the rest frequency 
of that transition for a cometocentric velocity of 0 km~s$^{-1}$ (dashed line). 
The CH$_3$OH detections reported by the 12m data are all E-type transitions.  Similarly, 
Figure 3 shows the spectrum of a cluster of E-type transitions of
CH$_3$OH centered around 157.3 GHz observed toward Comet LINEAR with the ARO 12m telescope.  

Finally, Figure 4 shows the spectrum taken at the frequency of the $J=8-7$ ($K$=$0$ through $7$) 
transitions of the symmetric top species methyl cyanide (CH$_3$CN) toward Comet Hale-Bopp using the 12m.  
The $K$=0 through 6 components are clearly present in these
data. CH$_3$CN and other symmetric tops have 
properties that make them ideal probes of physical conditions of
astronomical environments 
(see e.g.\ Araya et al.\ 2005; Remijan et al.\ 2004;  Hofner et al.\ 1996 ). 
As with CH$_3$OH, there is no apparent additional velocity components of CH$_3$CN 
or offset from the cometocentric rest velocity of 0 km s$^{-1}$.

Table 4 lists the molecular species that were searched for toward Comets Hale-Bopp and LINEAR at
or near the cometocentric velocity (0 km~s$^{-1}$).  Least square Gaussian fits were made for each 
spectral line in order to obtain the peak intensities and line-widths for the detected transitions. For
those transitions not detected, the 1 $\sigma$ rms noise level is given. The spectroscopic parameters
used in determining the column densities and production rates are found in Table 3 and the formalism 
for calculating column densities and production rates is described in $\S$4.1.

\section{ANALYSIS AND DISCUSSION}

\subsection{EXCITATION, ABUNDANCES AND PRODUCTION RATES}

Determining the best temperature, column density, and production rates of cometary molecules
are important because they are essential to understanding the formation of cometary species.  This includes
the role of grains in molecule formation, the relative impact of warm
versus cold gas phase chemistry, or the importance of
photodissociation in the production of daughter species in the coma.
In order to determine the temperatures and column 
densities of the region of the coma that contains the CH$_3$OH and CH$_3$CN emission, we assume that each region has 
uniform physical conditions, that the populations of the energy levels can be characterized by 
a Boltzmann distribution and finally, that the emission is optically thin. The 
rotational temperature diagram method was not employed for this analysis because even small errors in the intensities and 
widths of inherently weak spectral lines can cause large errors in determining the temperature and total 
column density (see e.g. Snyder et al.\ 2005).  The complete formalism for these
calculations using the BIMA array data can be found in Remijan et al.\ (2006) and for the
12m data, in Milam et al.\ (2006).

Figure 1 shows the results of the model fit (blue trace) applied to the CH$_3$OH BIMA array data.  The 
array data are well fit by an excitation temperature of T$_{ex}$=105$\pm$5 K and a total beam-averaged 
column density of N$_{T}$=$9\pm1\times10^{14}$~cm$^{-2}$ assuming that the CH$_3$OH emission is extended beyond
the 10.$''$33$\times$8.$''$46 average synthesized beam of the BIMA array (i.e.\ the beam filling factor, f=1).  
Similarly, Figure 2 shows the results of the model fit applied to the CH$_3$OH ARO 12m data with the same 
excitation temperature and a slightly lower column density of N$_{T}$=$2.4\pm0.3\times10^{14}$~cm$^{-2}$.  
The average total column density from these two facilities is N$_{T}$=$5.7\pm0.7\times10^{14}$~cm$^{-2}$.
As the figure shows, the modeling reproduces the observations.
Assuming CH$_3$OH is a parent species, production rates were determined from a Monte Carlo model 
for a nuclear source (see Combi \& Smyth, 1988).  This model traces the trajectories of molecules, 
within the telescope beam, ejected from the comet surface.  The observed column density is then 
matched for an output molecular production rate, Q (s$^{-1}$).   Input parameters include the lifetime of 
the molecule, from Huebner et al. (1992), scaled by R$_h^2$, the outflow velocity, telescope beam size, and 
the observed column density.  Assuming a nuclear origin for CH$_3$OH and using this Monte Carlo model, 
we find an average CH$_3$OH production rate from both instruments of $Q(CH_3OH)$$\sim$1.3$\times$10$^{29}$ 
s$^{-1}$ or $Q(CH_3OH)$/$Q(H_2O)$$\sim$1.3\% at $\sim$1 AU.  This production rate ratio assumes a  
$Q(H_2O)$=10$^{31}$ s$^{-1}$ (Crovisier et al.\ 2004).

In Figure 2, there is a slight discrepancy between the observed intensities and model predictions for the 
CH$_3$OH lines at 151.86 GHz. The model predicts a higher line intensity than observed and the model underestimates the
line intensity at 148.11 GHz and 145.76 GHz.  However, given the S/N ratio of all the passbands where these
lines are observed, the model does a very good job in predicting the expected line intensity of each 
CH$_3$OH transition observed with the 12m telescope.  The same can be said for the array
data.  The only discrepancy is the 96.7445 GHz,
J(K$_a$,K$_c$)=2(0,2)-1(0,1) E CH$_3$OH transition (Figure 1); 
the intensity is somewhat underestimated by our model fit.

Figure 4 shows the best model fit to the $J(K)=8(K)-7(K)$ ($K$=$0$ through $6$) transitions of methyl cyanide 
(CH$_3$CN) toward Comet Hale-Bopp. Once again, the blue trace shows the model predictions for the CH$_3$CN data 
which is best fit by an excitation temperature, T$_{ex}$=200$\pm$10 K
and a total beam-averaged column density, N$_{T}$=$2.6\pm0.3\times10^{12}$~cm$^{-2}$.  In this case, 
the predicted $K=7$ line intensity is below the 1 $\sigma$ detection limit, which indicates that the emission feature 
near the $K=7$ rest frequency may not be real.  Assuming a nuclear origin for CH$_3$CN and 
using a Monte Carlo model, we find a production rate of $Q(CH_3CN)$$\sim$1.7$\times$10$^{27}$ s$^{-1}$ or 
$Q(CH_3CN)$/$Q(H_2O)$$\sim$0.017\% at $\sim$1 AU. This production rate is similar to what
Biver et al.\ (1997) found, $\sim$10$^{27}$ s$^{-1}$ at 1 AU, in their study. 

These 12 m data overlap in frequency with the (CH$_2$OH)$_2$ IRAM 30m study of Crovisier et al.\ (2004).  In the IRAM dataset, 
the claimed detection of the J(K$_a$,K$_c$,$v$)=15(7,9,0)-14(7,8,1)
and 15(7,8,0)-14(7,7,1) transitions at 147.13 GHz are nearly
coincident in frequency and show up as one feature given 
the spectral resolution.  Moreover, the (CH$_2$OH)$_2$ transitions
have an overall line intensity equal to that of the $J(K)=8(4)-7(4)$ transition of CH$_3$CN detected with the 
IRAM 30m.  From the ARO 12m data, we identify the same spectral features of CH$_3$CN
but do not detect any emission from the two transitions of (CH$_2$OH)$_2$ below the 1$\sigma$ limit.  
From the CH$_3$CN, CH$_3$OH and (CH$_2$OH)$_2$ 12m data, it is clear we are probing different physical environments 
in the coma of Comet Hale-Bopp.  A discussion of the distribution of these molecular species is in $\S$4.2.

Figure 3 shows the best model fit (blue trace) to the cluster of E-type transitions of CH$_3$OH 
centered around 157.26 GHz toward Comet LINEAR with the ARO 12m telescope.  There have been very few molecular line 
observations toward Comet LINEAR.  The most complete study of molecular species toward this object was by 
Remijan et al.\ (2006) using the BIMA array.  These authors detected the J(K$_a$,K$_c$)=3(1,3)-4(0,4) 
A$^+$ transition of CH$_3$OH with a synthesized beam size of $22.''3\times16.''9$.  Using the proposed
excitation temperature of 115 K (DiSanti et al.\ 2004; Magee-Sauer et al.\ 2004; K{\"u}ppers et al.\ 2004), 
Remijan et al.\ determined a total beam-averaged column density of CH$_3$OH of
N$_{T}$=$1.4\pm0.3\times10^{14}$~cm$^{-2}$.  However, as shown in figure 3, the 12m CH$_3$OH data 
are best fit by an excitation temperature of T$_{ex}$=35$\pm$5 K and a total beam-averaged column density of
N$_{T}$=$2.2\pm0.3\times10^{13}$~cm$^{-2}$ (blue trace), implying a production rate of 
$Q(CH_3OH)$$\sim$2.0$\times$10$^{27}$ s$^{-1}$ or $Q(CH_3OH)$/$Q(H_2O)$$\sim$1.5\% at 0.3 AU.  This production 
rate ratio assumes a $Q(H_2O)$=1.3$\times$10$^{29}$ s$^{-1}$ (Lecacheux et al.\ 2004).
To reconcile the differing column density determined by Remijan et al.\ (2006), 
we used the best model fit parameters of temperature and column
density from the 12m data to predict the line intensity of the J(K$_a$,K$_c$)=3(1,3)-4(0,4) A$^+$ transition
detected by the BIMA array. Figure 5 shows the result of this analysis (5(a) shows the hanning
smoothed data and 5(b), the raw data as shown in figure 1 of Remijan
et al. (2006)).  The fit is excellent.   From using the ARO data it is quite clear that
the assumption of an excitation temperature of T$_{ex}$=115 K for CH$_3$OH was incorrect and the determined
column density of CH$_3$OH toward Comet LINEAR was therefore overestimated by almost an order of
magnitude.

\subsection{DISTRIBUTION OF MOLECULAR SPECIES}

We present two complementary sets of data from Comet Hale-Bopp that give great insights into the distribution
of CH$_3$OH, CH$_3$CN and (CH$_2$OH)$_2$.  First, it is quite clear from the data presented in Figures 1
and 2 that despite the drastic change in the telescope beam sizes between the array and the 12m, we
are still sampling a region of the coma that is thermalized to $\sim$105 K.  The ratio of areas subtended by the 12m beam, 
compared to the BIMA array synthesized beam, is $\sim$21:1 which corresponds to a change in the physical sampled size 
scale from D$\sim$40000 km to D$\sim$8500 km. A direct comparison of the column densities from the 12m and BIMA array 
suggest an enhancement of the CH$_3$OH abundance at smaller coma
radii.  Furthermore, Column 5 in table 3 lists the Einstein A
coefficients for each transition of CH$_3$OH, CH$_3$CN and (CH$_2$OH)$_2$.  Assuming a
collision cross section of $\sim$10$^{-15}$ cm$^2$ and a temperature
of 105 K, the calculated critical densities of the detected CH$_3$OH
transitions range from (0.7-3.2)$\times$10$^5$ cm$^{-3}$. Our
observations encompass a region with a H$_2$O density of
$\sim$10$^{(5.7-7.5)}$ cm$^{-3}$ (Lovell et al.\ 2004).
Therefore, there is sufficient collisional excitation to populate the
CH$_3$OH levels, and our observations are sampling molecular abundances and not excitation effects.

Figure 6(a) once again shows the distribution of the J(K$_a$,K$_c$)=3(1,3)-4(0,4) A$^+$ transition of  CH$_3$OH around 
Comet Hale-Bopp.  Figure 6(b) is a model fit to the data shown in (a).  The model was generated assuming a spherical Haser model 
of parent species sublimating off the comet's nucleus.  From the data in (a), the scale length of the distribution of CH$_3$OH around
the nucleus is measured to be r$\sim$10$''$.  Using this scale length and the MIRIAD task IMGEN, a Haser model distribution of CH$_3$OH was
generated which was then convolved with the synthesized beam of the BIMA array (shown at the bottom left of each panel
in figure 6).  The residuals obtained by subtracting the model from the data are shown in figure 6(c).  While it appears 
that there may be a more extended distribution of CH$_3$OH detected by the BIMA array based on the residual data, it 
is important to note that the contour levels shown in (a) and (b) are 3, 4, 5 and 6 $\sigma$ ($\sigma$=0.05 Jy beam$^{-1}$), 
whereas the data shown in (c) are 1 and 1.5 $\sigma$. Although there is some indirect evidence for an extended distribution 
of CH$_3$OH with D = 40000 km from the 12 m detections of CH$_3$OH, the data presented in (c) indicate that any extended emission is 
is completely resolved out by the array.  The array would be sensitive to extended emission of 
CH$_3$OH in the outer coma only if the distribution was ``clumpy'', as is seen in H$_2$CO (Milam et al.\ 2006).  

It appears that the CH$_3$CN is tracing higher temperature and higher
density gas than CH$_3$OH.  Because CH$_3$CN has a larger dipole moment
than CH$_3$OH, its measured excitation should be lower, given identical conditions.
However, if the overall distribution of CH$_3$CN
is closer to the nucleus than CH$_3$OH, we would expect to measure a higher excitation temperature (it is well known that 
the temperature of the coma increases with decreasing coma radii (Biver et al.\ 1999)).  This distribution would 
also imply that CH$_3$CN is a parent species ($\S$4.3). Given the
difference in beam sizes between the 12m and 30m telescopes, we can get an estimate of the size of the
emitting region based on the relative beam sizes using the beam filling factor in the expression for column density. 
Comparing the relative line intensities between the 
$K$=3 and 4 transitions of CH$_3$CN from the IRAM 30m data and what is measured from the 12m, the 30m emission 
features from CH$_3$CN are $\sim$5-10$\times$ stronger than what is observed from the 12m.  This difference implies 
either 1) an enhancement of the CH$_3$CN column density at smaller
coma radii or 2) the excitation of the transitions we 
observed of CH$_3$CN takes place at a slightly smaller radii where the density is higher and the measurements taken 
with the 12m are suffering from some beam dilution.  Assuming the
H$_2$O density profile of Lovell et al.\ (2004), the maximum coma
radius that lies below the CH$_3$CN critical density range of (1.2-5.3)$\times$10$^6$ cm$^{-3}$ is $\sim$10000 km.
This explains why the 30m CH$_3$CN line strengths are higher than
those from the 12m.  Finally, from the non-detection of 
(CH$_2$OH)$_2$ in the 12m observations, it is also apparent that the distribution of (CH$_2$OH)$_2$ must also be 
compact in the cometary coma.  Assuming the detected transition line
strength of $\sim$0.09 K in the 30m spectrum is from (CH$_2$OH)$_2$ and a 
1$\sigma$ line intensity of $\sim$0.01 K in the 12m passband, the emission from (CH$_2$OH)$_2$ must be on the order of 
D$<$9$''$ corresponding to a physical size of D$<$8500 km.  High sensitivity, high resolution interferometer 
observations are necessary to confirm this distribution. 

\subsection{FORMATION OF COMETARY METHANOL AND OTHER MOLECULAR SPECIES}

From the high resolution methanol observations
of the BIMA array, it is clear that the CH$_3$OH is either sublimating directly off the cometary ices contained in the
nucleus or is formed very deep in cometary coma.  If CH$_3$OH 
originates in the ices, this molecules may
be a remnant from the formation of the presolar nebula and hence the
interstellar medium. Currently, there are two accepted formation pathways
to the interstellar production of CH$_3$OH, including the radiative association of CH$_3^+$ + H$_2$O $\rightarrow$ CH$_3$OH$_2^+$, followed
by recombination with an electron to produce CH$_3$OH and H (Herbst,
1985). Grain surface reactions that involve the repeated hydrogenation
of CO $\rightarrow$ H + CO $\rightarrow$ HCO + H $\rightarrow$ HCHO
$\rightarrow$ HCHOH are thought to eventually lead to CH$_3$OH (Hiraoka 
et al.\ 1994).  Our  observations show there is no significant
enhancement in the production of CH$_3$OH in the outer coma as the comet enters the inner
solar system.  

The same argument can be applied to CH$_3$CN.  The accepted route to the formation of CH$_3$CN in interstellar
environments is CH$_3^+$ + HCN, a collision complex is formed that equilibrates to CH$_3$CNH$^+$ + $\nu$ (Herbst, 1985); then
CH$_3$CNH$^+$ combines with an electron to form CH$_3$CN + H. Our
observations of CH$_3$CN show that it is present close to the cometary
nucleus, indicating direct sublimation off cometary ices. However, the excitation of the
transitions we observed of CH$_3$CN takes place at a smaller radii
where the density is higher, and this molecule may be present as well at larger radii.
However, the true spatial scale and scale length of CH$_3$CN needs to be verified by interferometric observations.

Finally, the non-detection of (CH$_2$OH)$_2$ with the ARO 12m gives insight into the formation of (CH$_2$OH)$_2$
as well as the possibility of detecting CH$_2$OHCHO.  If the
transition detected by the 30m is indeed from cometary
 (CH$_2$OH)$_2$, in $\S$4.2, we found the predicted distribution of (CH$_2$OH)$_2$ toward Comet
Hale-Bopp was calculated to be  D$<$9$''$ ($<$8500 km), again
suggesting direct sublimation off cometary ices or 
an enhanced production in the inner coma.  
Presumably, as with the formation of CH$_3$OH and CH$_3$CN in interstellar environments, 
(CH$_2$OH)$_2$ would be formed in the ISM and then seeded into cometary ices.  Currently, there is no accepted gas phase 
interstellar formation mechanism that can lead to the production of (CH$_2$OH)$_2$.  However, there are several formation 
pathways using surface chemistry which may occur on icy grain mantles.
For example, to form (CH$_2$OH)$_2$ and CH$_2$OHCHO, formaldehyde
(H$_2$CO) produces CH$_2$OHCHO in an aqueous 
Formose reaction, which can then be hydrogenated 
to form (CH$_2$OH)$_2$ (e.g., see Walker 1975).  Charnley (2001) predicts that (CH$_2$OH)$_2$ and CH$_2$OHCHO 
could be formed by direct hydrogenation reactions, starting from ketene (CH$_2$CO) on grain surfaces. 
Finally, Hudson \& Moore (2000) showed that proton irradiated CH$_3$OH on icy interstellar grain mantles can lead to the formation of 
(CH$_2$OH)$_2$.  Thus, there appear to be several ways to form and
embed (CH$_2$OH)$_2$ into cometary ices.


\section{CONCLUSIONS}

We presented an interferometric and single dish study of cometary molecules toward comets
C/1995 O1 (Hale-Bopp) and C/2002 T7 (LINEAR) using the Berkeley-Illinois-Maryland Association (BIMA) interferometer 
at 3 mm and the Arizona Radio Observatory (ARO) 12m telescope at 2 mm.  The overall conclusions from our analysis
of these data are:

\begin{enumerate}
 
\item The CH$_3$OH Hale-Bopp data are well fit by an excitation temperature of T$_{ex}$=105$\pm$5 K and a total beam-averaged 
column density of N$_{T}$=$5.7\pm0.7\times10^{14}$~cm$^{-2}$.  Assuming a nuclear origin for CH$_3$OH and using a Monte Carlo 
model, we find an average CH$_3$OH production rate from both instruments of $Q(CH_3OH)$$\sim$1.3$\times$10$^{29}$ s$^{-1}$ 
or $Q(CH_3OH)$/$Q(H_2O)$$\sim$1.3\% at $\sim$ 1 AU. 

\item The CH$_3$OH ARO 12m LINEAR data centered around 157.26 GHz are best fit by an excitation temperature of 
T$_{ex}$=35$\pm$5 K and a total beam-averaged column density of N$_{T}$=2.2$\pm$0.3$\times10^{13}$~cm$^{-2}$ and thus, a production 
rate $Q(CH_3OH)$$\sim$2.0$\times$10$^{27}$ s$^{-1}$ or $Q(CH_3OH)$/$Q(H_2O)$=1.5\% at 0.3 AU.  

\item  From the combination of the single-dish and aperture synthesis observations of CH$_3$OH, we
find the distribution of CH$_3$OH toward Comet Hale-Bopp is well described by a spherical outflow with an increase
in column density closer to the cometary nucleus.  The 
data presented in the array images show no evidence of significant enhanced production of CH$_3$OH in the extended coma
or from jets as any extended emission is completely resolved out by the array.

\item The CH$_3$CN 12m Hale-Bopp data is best fit by an excitation temperature, T$_{ex}$=200$\pm$10 K 
and a total beam-averaged column density, N$_{T}$=2.6$\pm$0.3$\times10^{12}$~cm$^{-2}$. Assuming a nuclear origin 
for CH$_3$CN and using a Monte Carlo model, we find an average CH$_3$CN production rate of $Q(CH_3CN)$$\sim$1.7$\times$10$^{27}$ s$^{-1}$ 
or $Q(CH_3CN)$/$Q(H_2O)$$\sim$0.017\% at $\sim$ 1 AU.  A comparison between the single-dish observations from the
ARO 12m and the IRAM 30m of CH$_3$CN suggest that the ARO observations
are beam diluted.  The excitation of the transitions observed of CH$_3$CN takes place at a smaller radii where the density
is higher and suggest a nuclear origin of CH$_3$CN. 

\item The non-detection of a previously claimed transition of cometary (CH$_2$OH)$_2$ toward Comet 
Hale-Bopp with the ARO 12m telescope indicates a compact distribution of emission on the order of 
$<$9$''$ ($<$8500 km).  This supports the hypothesis that the cometary production of (CH$_2$OH)$_2$
is direct sublimation off cometary ices from the nucleus.  


We thank D. K. Yeomans for ephemerides assistance and G. Engargiola, T. Helfer, W. Hoffman, R. L. Plambeck, and 
M. W. Pound for invaluable technical contributions. We also thank an anonymous referee for a favorable review of this
work and whose comments and suggestions provided additional clarity to this manuscript.  This material is based on 
work supported by the National 
Aeronautics and Space Administration through the NASA Astrobiology Institute under Cooperative Agreement 
CAN-02-OSS-02 issued through the Office of Space Science. S. N. M. would like to thank the Phoenix Chapter of 
ARCS, specifically Mrs. Scott L. Libby, Jr. endowment, for partial funding. This work was partially funded by 
NASA NAG5-4292, NAG5-4080, NAG5-8708, and NGT5-0083; NSF AST 96-13998, AST96-13999, AST96-13716, AST96-15608, and 
AST99-81363; and the Universities of Illinois, Maryland, and California, Berkeley. M. W. was funded by NSF AST 
9625360, AST 9796263, and AST 0098583 and NASA NAG5-4349.

\end{enumerate}

\clearpage
\begin{deluxetable}{cccccc}
\tablecolumns{6}
\scriptsize
\tablecaption{CH$_3$OH Observational Parameters - BIMA array}
\tablewidth{400pt}
\tablehead{
\colhead{Frequency} & \colhead{Observation} & \colhead{$\Delta$} & \colhead{r} & \colhead{beam} & \colhead{chan rms}\\
\colhead{(GHz)} & \colhead{date} & \colhead{(AU)} & \colhead{(AU)} & \colhead{($''$$\times$$''$)} & \colhead{(Jy beam$^{-1}$)}
}
\startdata
96.741 & 1997 Mar 27 & 1.323 & 0.919 & 10.8$\times$8.2 & 0.21\\
       & 1997 Mar 31 & 1.344 & 0.914 & 9.8$\times$7.8 & 0.20\\
107.014 & 1997 Mar 26 & 1.320 & 0.921 & 19.3$\times$6.8 & 0.43\\
        & 1997 Apr 02 & 1.359 & 0.914 & 11.7$\times$6.6 & 0.15\\
        & 1997 Apr 03 & 1.367 & 0.915 & 16.0$\times$6.4 & 0.30\\
\enddata
\end{deluxetable}

\clearpage
\begin{deluxetable}{cccccccc}
\tablecolumns{8}
\scriptsize
\tablecaption{Observational Parameters - ARO 12m}
\tablewidth{400pt}
\tablehead{
\colhead{Frequency} & \colhead{Observation} & \colhead{$\Delta$} & \colhead{r} & \colhead{beam} & \colhead{D} & \colhead{$\eta_c$} & \colhead{molecule}\\
\colhead{(GHz)} & \colhead{date} & \colhead{(AU)} & \colhead{(AU)} & \colhead{($''$)} & \colhead{(km)} & \colhead{} & \colhead{}
}
\startdata
\multicolumn{7}{c}{Comet Hale-Bopp}\\
\hline
145.766 & 1997 Mar 20.94 & 1.318 & 0.940 & 43 & 41105 & 0.76 & CH$_3$OH \\
147.105\tablenotemark{a} & 1997 Mar 20.66 & 1.318 & 0.940 & 43 & 41105 & 0.76 & CH$_3$CN \\
148.112 & 1997 Mar 11.89 & 1.368 & 0.989 & 42 & 41672 & 0.76 & CH$_3$OH \\
150.142 & 1997 Mar 10.71 & 1.377 & 0.996 & 42 & 41946 & 0.75 & CH$_3$OH \\
151.860 & 1997 Mar 20.92 & 1.318 & 0.940 & 41 & 39193 & 0.74 & CH$_3$OH \\
157.049 & 1997 Mar 21.54 & 1.316 & 0.936 & 40 & 38179 & 0.72 & CH$_3$OH \\
\multicolumn{7}{c}{Comet LINEAR}\\
\hline
157.261\tablenotemark{b} & 2004 May 21.80 & 0.284 & 0.865 & 40 & 8239 & 0.72 & CH$_3$OH \\
\enddata
\tablenotetext{a}{Average of observing frequencies between the $K$=7 and $K$=0 lines of CH$_3$CN.} 
\tablenotetext{b}{Average of observing frequencies of CH$_3$OH lines detected in the same spectral passband.} 
\end{deluxetable}

\clearpage
\begin{deluxetable}{lrcccc}
\tablecolumns{6}
\tablecaption{Molecular Line Parameters}
\tablewidth{500pt}
\tablehead{
\colhead{Molecule} & \colhead{Transition} & \colhead{Frequency} & \colhead{$<S_{i,j}\mu^{2}>$} & \colhead{$A_{i,j}$} & \colhead{E$_{u}$}\\
\colhead{} & \colhead{} & \colhead{(MHz)} & \colhead{(Debye$^{2}$)} & \colhead{(s$^{-1}$)} & \colhead{(K)}
}
\startdata
CH$_{3}$OH\tablenotemark{a} & J(K$_a$,K$_c$)=2(-1,2)-1(-1,1) E & 96,739.363(3) & 1.2 & 2.55$\times$10$^{-6}$ & 12.6\\
 & 2(0,2)-1(0,1) A$^+$ & 96,741.377(3) & 1.6 & 3.41$\times$10$^{-6}$ & 7.0\\
 & 2(0,2)-1(0,1) E & 96,744.549(3) & 1.6 & 3.41$\times$10$^{-6}$ & 20.1\\
 & 2(1,1)-1(1,0) E & 96,755.507(3) & 1.2 & 2.62$\times$10$^{-6}$ & 28.0\\
 & 3(1,3)-4(0,4) A$^+$ & 107,013.770(13) & 3.0 & 6.13$\times$10$^{-6}$& 28.4\\
 & 16(0,16)-16(-1,16) E & 145,766.163(27) & 7.0 &7.60$\times$10$^{-6}$ & 327.9\\
 & 15(0,15)-15(-1,15) E & 148,111.919(24) & 7.3 &8.80$\times$10$^{-6}$ & 290.9\\
 & 14(0,14)-14(-1,14) E & 150,141.593(22) & 7.5 &10.2$\times$10$^{-6}$ & 256.3\\
 & 13(0,13)-13(-1,13) E & 151,860.170(20) & 7.6 &11.5$\times$10$^{-6}$ & 223.9\\
 & 6(0,6)-6(-1,6) E & 157,048.586(13) & 5.7 &1.96$\times$10$^{-6}$ & 61.9\\
 & 4(0,4)-4(-1,4) E & 157,246.041(14) & 4.2 &21.0$\times$10$^{-6}$ & 36.4\\
 & 1(0,1)-1(-1,1) E & 157,270.818(15) & 1.5 &22.1$\times$10$^{-6}$ & 15.5\\
 & 3(0,3)-3(-1,3) E & 157,272.320(14) & 3.3 &21.5$\times$10$^{-6}$ & 27.1\\
 & 2(0,2)-2(-1,2) E & 157,276.004(14) & 2.4 &21.8$\times$10$^{-6}$ & 20.1\\
 &  &  &  & \\
CH$_{3}$CN\tablenotemark{a} & J(K)=8(6)-7(6) & 147,072.608(2) & 53.2 & 0.58$\times$10$^{-4}$ & 289.0\\
 & 8(5)-7(5) & 147,103.741(1) & 74.2 & 1.62$\times$10$^{-4}$ & 210.5\\
 & 8(4)-7(4) & 147,129.232(1) & 91.3 & 1.99$\times$10$^{-4}$ & 146.2\\
 & 8(3)-7(3) & 147,149.068(1) & 104.6 & 1.14$\times$10$^{-4}$ & 96.1\\
 & 8(2)-7(2) & 147,163.243(1) & 114.1 & 2.49$\times$10$^{-4}$ & 60.4\\
 & 8(1)-7(1) & 147,171.750(1) & 119.8 & 2.61$\times$10$^{-4}$ & 38.9\\
 & 8(0)-7(0) & 147,174.587(1) & 121.7 & 2.66$\times$10$^{-4}$ & 31.8\\
 &  &  &  & \\
(CH$_{2}$OH)$_{2}$\tablenotemark{b} &  J(K$_a$,K$_c$,$v$)=15(7,9,0)-14(7,8,1) & 147,131.814(36) & 2.5 & 5.99$\times$10$^{-5}$ & 83.1\\
 & 15(7,8,0)-14(7,7,1) & 147,132.412(36) & 2.7 & 5.99$\times$10$^{-5}$
& 83.1\\
\enddata
\tablenotetext{a}{Molecular line parameters of CH$_3$OH and CH$_3$CN taken from SLAIM at www.splatalogue.net.  
Q$_{r}$=1.2T$_{r}^{\frac{3}{2}}$ for CH$_3$OH and Q$_{r}$=3.9T$_{r}^{\frac{3}{2}}$ for CH$_3$CN.} 
\tablenotetext{b}{Molecular line parameters of (CH$_{2}$OH)$_{2}$
  taken from CDMS (M{\"u}ller et al.\ 2005).  The energy levels are
  noted as J(K$_a$,K$_c$,$v$) where $v$ is the quantum number
  associated with OH tunneling (Christen \& M{\"u}ller 2003).} 
\end{deluxetable}

\clearpage
\begin{deluxetable}{rcccccccc}
\tabletypesize{\tiny}
\tablewidth{40pc}
\tablecolumns{9}
\tablecaption{Comet Hale-Bopp and Comet LINEAR Molecular Line Identifications}
\tablehead{
\colhead{Species} & \colhead{Telescope} & \colhead{$\nu$} &
\colhead{Transition} & \colhead{$I_{line}$\tablenotemark{a}} & \colhead{$\Delta v$} & \colhead{N$_{T}$} & \colhead{Q$_{p}$} & \colhead{Q$_{X}$/Q$_{H_2O}$}\\
\colhead{} & \colhead{} & \colhead{(MHz)} & \colhead{} & \colhead{} & \colhead{(km s$^{-1}$)} & \colhead{(cm$^{-2}$)} & \colhead{(s$^{-1}$)}  & \colhead{}
}
\startdata
\multicolumn{9}{c}{Comet Hale-Bopp}\\
\hline
CH$_3$OH & BIMA & 96,739.4 &  2(-1,2)-1(-1,1) E & 0.31(5) & 2.4(2) & 9$\pm$1$\times$10$^{14}$ & 1.1$\pm$0.3$\times$10$^{29}$ & $\sim$1.1$\times$10$^{-2}$\tablenotemark{b} \\
 &  & 96,741.4 & 2(0,2)-1(0,1) A$^+$ & 0.52(5) & 2.0(2) & & & \\
 &  & 96,744.5 & 2(0,2)-1(0,1) E & 0.60(5) & 2.1(2) & & & \\
 &  & 96,755.5 & 2(1,1)-1(1,0) E & 0.29(5) & 3.2(2) & & &\\
 &  & 107,013.8 & 3(1,3)-4(0,4) A$^+$ & 1.12(5) & 1.9(2) & & &\\
CH$_3$OH & 12m & 145,766.2 & 16(0,16)-16(-1,16) E & 0.03(2) & 2.0(8) & 2.4$\pm$0.3$\times$10$^{14}$ & 1.5$\pm$0.3$\times$10$^{29}$ & $\sim$1.5$\times$10$^{-2}$ \\
 &  & 148,111.9 & 15(0,15)-15(-1,15) E & 0.04(2) & 3.0(8) & & & \\
 &  & 150,141.6 & 14(0,13)-14(-1,14) E & 0.04(1) & 1.9(8) & & & \\
 &  & 151,860.2 & 13(0,13)-13(-1,13) E & 0.05(1) & 2.5(8) & & & \\
 &  & 157,048.6 & 6(0,6)-6(-1,6) E & 0.23(1) & 2.4(8) & & & \\
& & & & & & & & \\
& & & & {\bf CH$_3$OH} & {\bf AVERAGE:} & 5.7$\pm$0.7$\times$10$^{14}$ &1.3$\pm$0.3$\times$10$^{29}$ & $\sim$1.3$\times$10$^{-2}$\\
& & & & & & & & \\
CH$_3$CN & 12m & 147,129.2 & 8(4)-7(4) & 0.019(1) & 1.6(8)  & 2.6$\pm$0.3$\times$10$^{12}$ & 1.7$\pm$0.3$\times$10$^{27}$ & $\sim$1.7$\times$10$^{-4}$ \\
 &  & 147,149.0 & 8(3)-7(3) & 0.054(1) & 2.8(8) & & & \\
 &  & 147,163.2 & 8(2)-7(2) & 0.034(1) & 3.2(8) & & & \\
 &  & 147,171.7 & 8(1)-7(1) & 0.049(1) & 2.4(8) & & & \\
 &  & 147,174.5 & 8(0)-7(0) & 0.051(1) & 2.4(8) & & & \\
& & & & & & & \\
(CH$_{2}$OH)$_{2}$ & 12m & 147,131.8 & 15(7,*,0)-14(7,*,0) & $<$0.01 & & & & \\
\hline
\multicolumn{9}{c}{Comet T7/LINEAR}\\
\hline
CH$_3$OH & 12m & 157,246.1 & 6(0,6)-6(-1,6) E & 0.087(1) & 1.9(10) & 2.2$\pm$0.3$\times$10$^{13}$ & 2.0$\pm$0.3$\times$10$^{27}$ & $\sim$1.5$\times$10$^{-2}$\tablenotemark{c}\\
 & & 157,270.9 & 1(0,1)-1(-1,1) E & 0.051(1) & 1,9(10) & & & \\
 & & 157,272.3 & 3(0,3)-3(-1,3) E & 0.101(1) & 1,9(10) & & & \\
 & & 157,276.0 & 2(0,2)-2(-1,2) E & 0.058(1) & 1,9(10) & & & \\
\enddata
\tablenotetext{a}{For the BIMA data, $I_{line}$ represents $\Delta I$
  and are in units of Jy beam$^{-1}$. For the 12m observations, $I_{line}$ represents T$_R^*$ in units of K.}
\tablenotetext{b}{Comet Hale-Bopp: $Q(H_2O)$=10$^{31}$s$^{-1}$ (Crovisier et al.\ 2004)}
\tablenotetext{c}{Comet T7/LINEAR: $Q(H_2O)$=1.3$\times$10$^{29}$s$^{-1}$ (Milam et al.\ 2006)}

\end{deluxetable}

\clearpage

\figcaption{CH$_3$OH lines detected toward Comet Hale-Bopp with the BIMA array.
(a) Emission contours from the 2(0,2)-1(0,1) E transition of CH$_3$OH at 96.741 GHz. 
Contours indicate the location of the 2(0,2)-1(0,1) E CH$_3$OH emission. The contour 
levels are -0.2, 0.4, 0.6, 0.8, 1.0, and 1.2 Jy beam$^{-1}$. The synthesized 
beam of $10\farcs 3\times8\farcs 5$ is indicated at the bottom left corner.
(b) CH$_3$OH spectrum toward Comet Hale-Bopp (Hanning smoothed over three channels).  The rms noise level is $\sim$0.1 Jy 
beam$^{-1}$ (indicated by the vertical bar at the left of the spectrum).  The spectral 
line labels correspond to the rest frequency located at the top left of the spectral 
window for a velocity of 0 km~s$^{-1}$.  The dashed line is centered on this velocity.  
The blue trace shows a model fit (see text).
(c) Emission contours from the 3(1,3)-4(0,4) A$^+$ transition of CH$_3$OH at 107.014 GHz. 
Contours indicate the location of the CH$_3$OH emission 
averaged between -2 km~s$^{-1}$ and 2 km~s$^{-1}$.  The contour 
levels are -0.05, 0.20, 0.30, 0.40, 0.50, and 0.60 Jy beam$^{-1}$. The synthesized 
beam of $10\farcs 8\times7\farcs 2$ is indicated at the bottom left corner.
(d) CH$_3$OH spectrum toward Comet Hale-Bopp.  The rms noise level is $\sim$0.1 Jy 
beam$^{-1}$ (indicated by the vertical bar at the left of the spectrum).  The spectral 
line label is the same as in (b) and the blue trace shows a model fit (see text).}

\figcaption{CH$_3$OH lines detected toward Comet Hale-Bopp with the 12m telescope.  The backends employed for 
the presented observations were either filter banks with 500 kHz resolution or a millimeter
autocorrelator (MAC) with a resolution of 781 kHz. The 
spectral line labels correspond to the rest frequency located at the top left of the spectral 
window for a velocity of 0 km~s$^{-1}$ (dashed line). The blue traces show a model fit (see text).}

\figcaption{CH$_3$OH lines detected toward Comet LINEAR with the 12m
  telescope.  Spectral resolution is 500 kHz and 
the spectral line labels are similar to Figure 2. The blue traces show a model fit (see text).}

\figcaption{CH$_3$CN lines detected toward Comet Hale-Bopp with the
  12m telescope. Spectral resolution is 781 kHz and the 
spectral line labels are similar to Figure 2. The blue trace shows the
model fit (see text).}

\figcaption{CH$_3$OH transitions from Comet LINEAR and the model fit to the
data (blue trace) from the BIMA array data.}

\figcaption{CH$_3$OH Haser model fit to the BIMA array data. (a) The emission contours from the 3(1,3)-4(0,4) A$^+$ transition of
  CH$_3$OH at 107.014 GHz. (b) The Haser model predictions to these data assuming a scale length of
  r$\sim$10$''$ and then convolved with the synthesized beam (shown at
  the bottom left).  (c) The residuals obtained 
by subtracting the model from the data.}

\clearpage
\begin{figure}
\epsscale{0.8}
\plotone{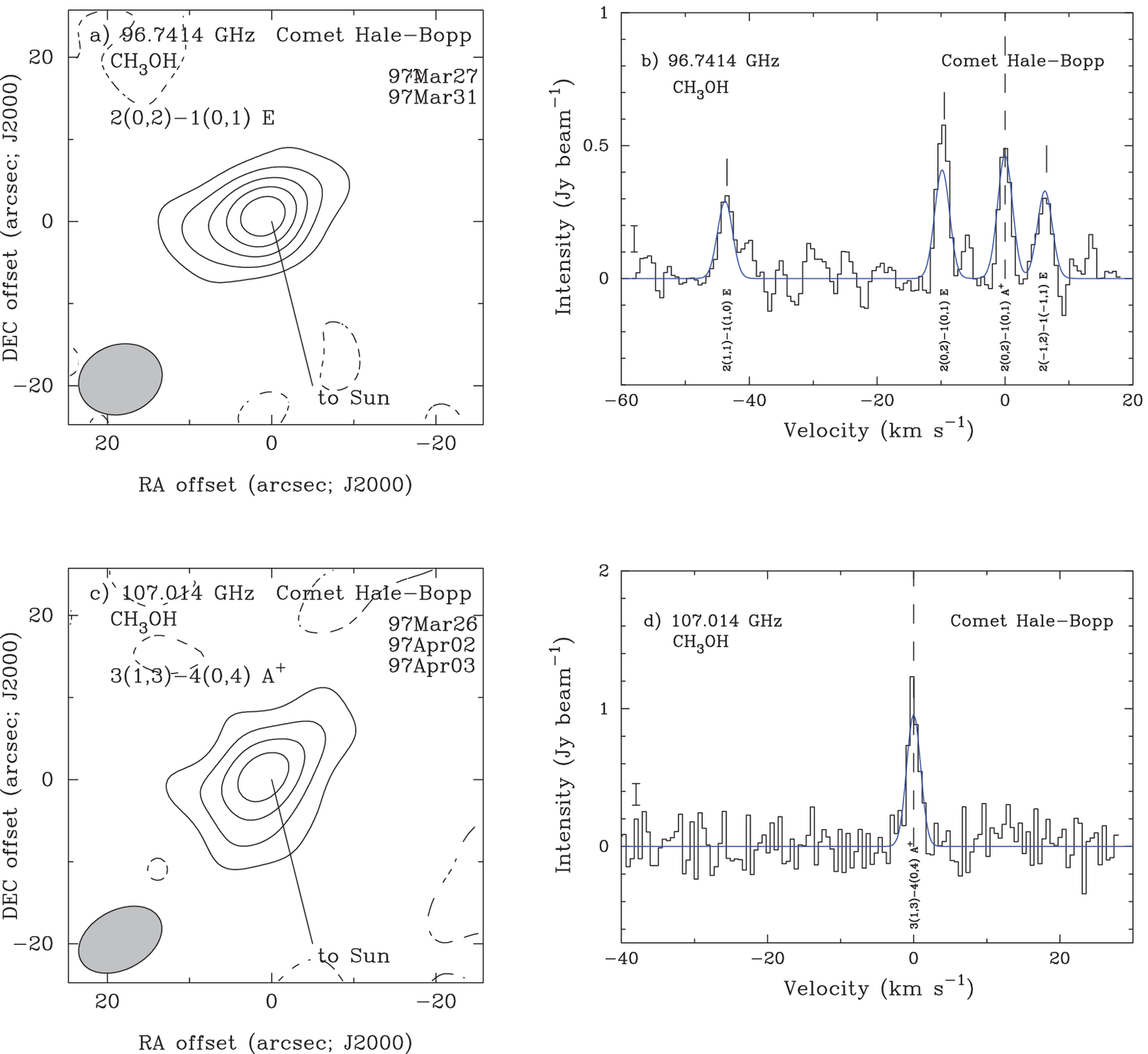}
\centerline{Figure 1.}
\end{figure}
\clearpage
\begin{figure}
\epsscale{0.5}
\plotone{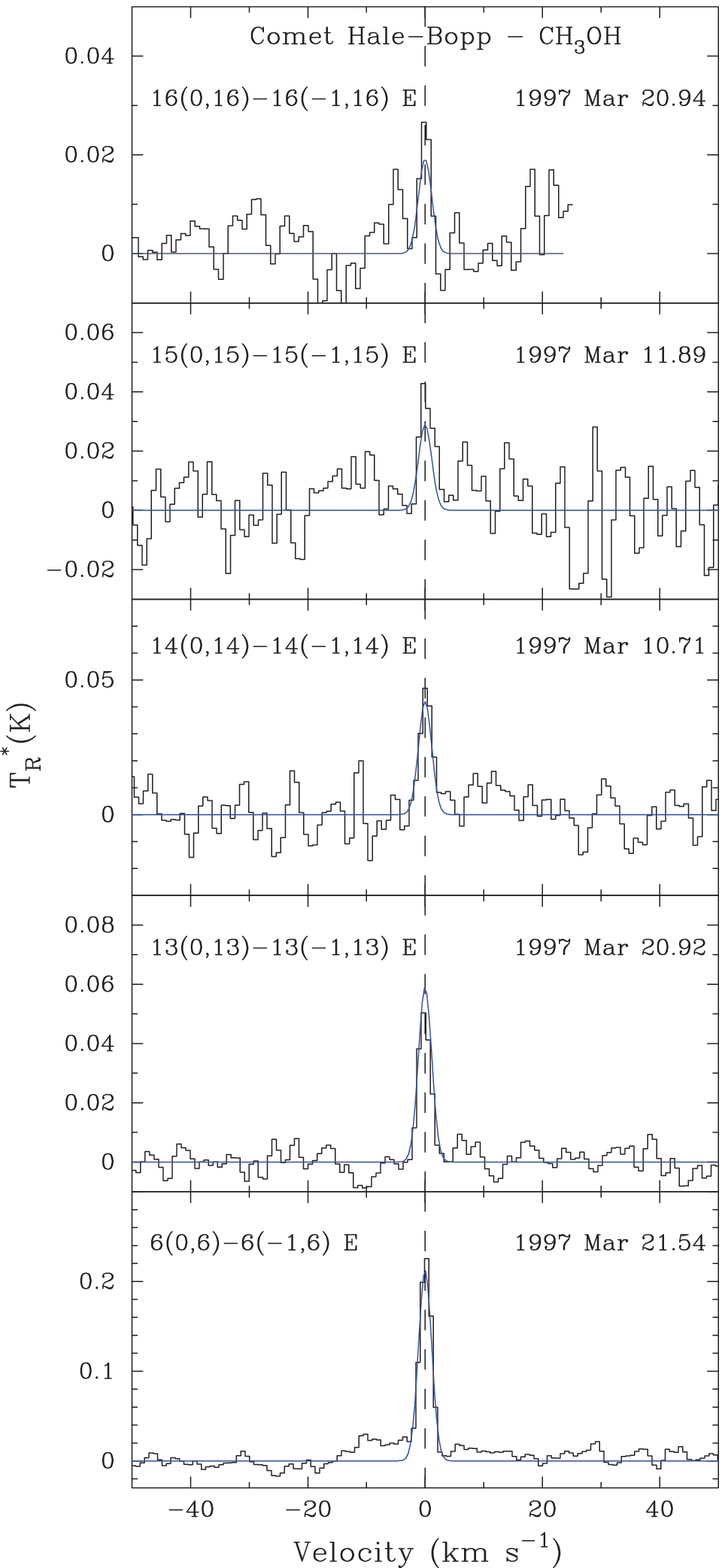}
\centerline{Figure 2.}
\end{figure}
\clearpage
\begin{figure}
\epsscale{0.8}
\plotone{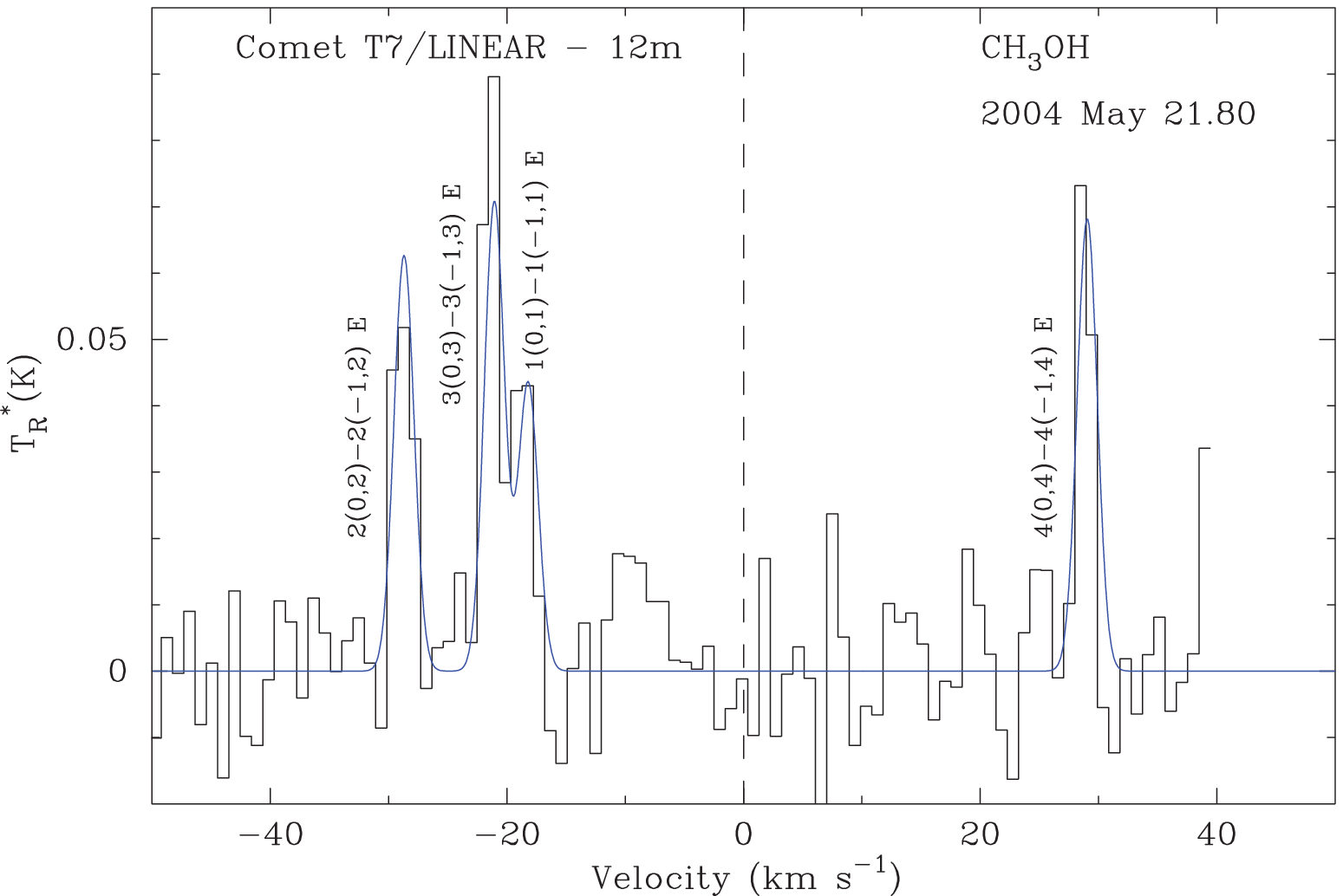}
\centerline{Figure 3.}
\end{figure}
\clearpage
\begin{figure}
\epsscale{0.8}
\plotone{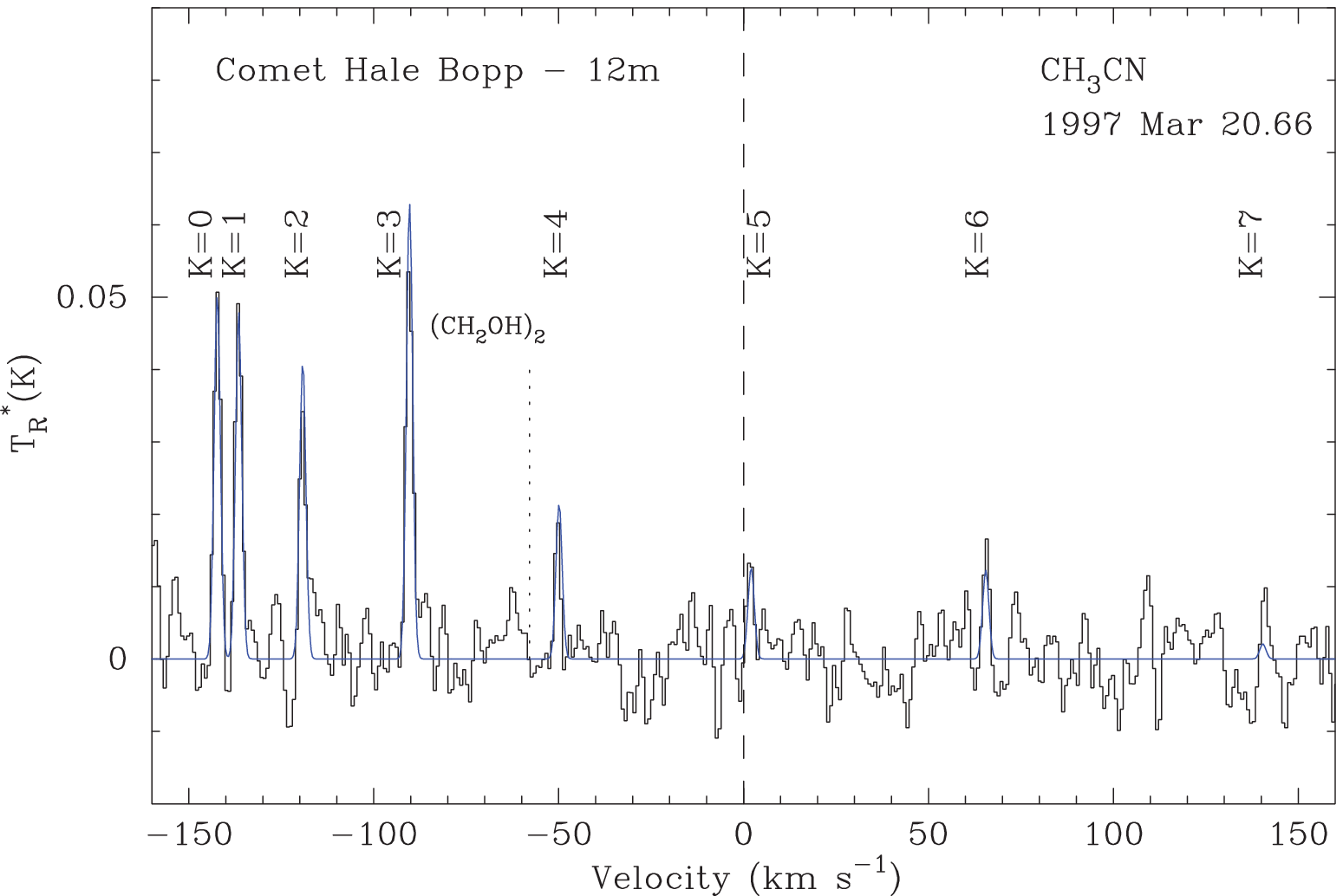}
\centerline{Figure 4.}
\end{figure}
\clearpage
\clearpage
\begin{figure}
\epsscale{0.8}
\plotone{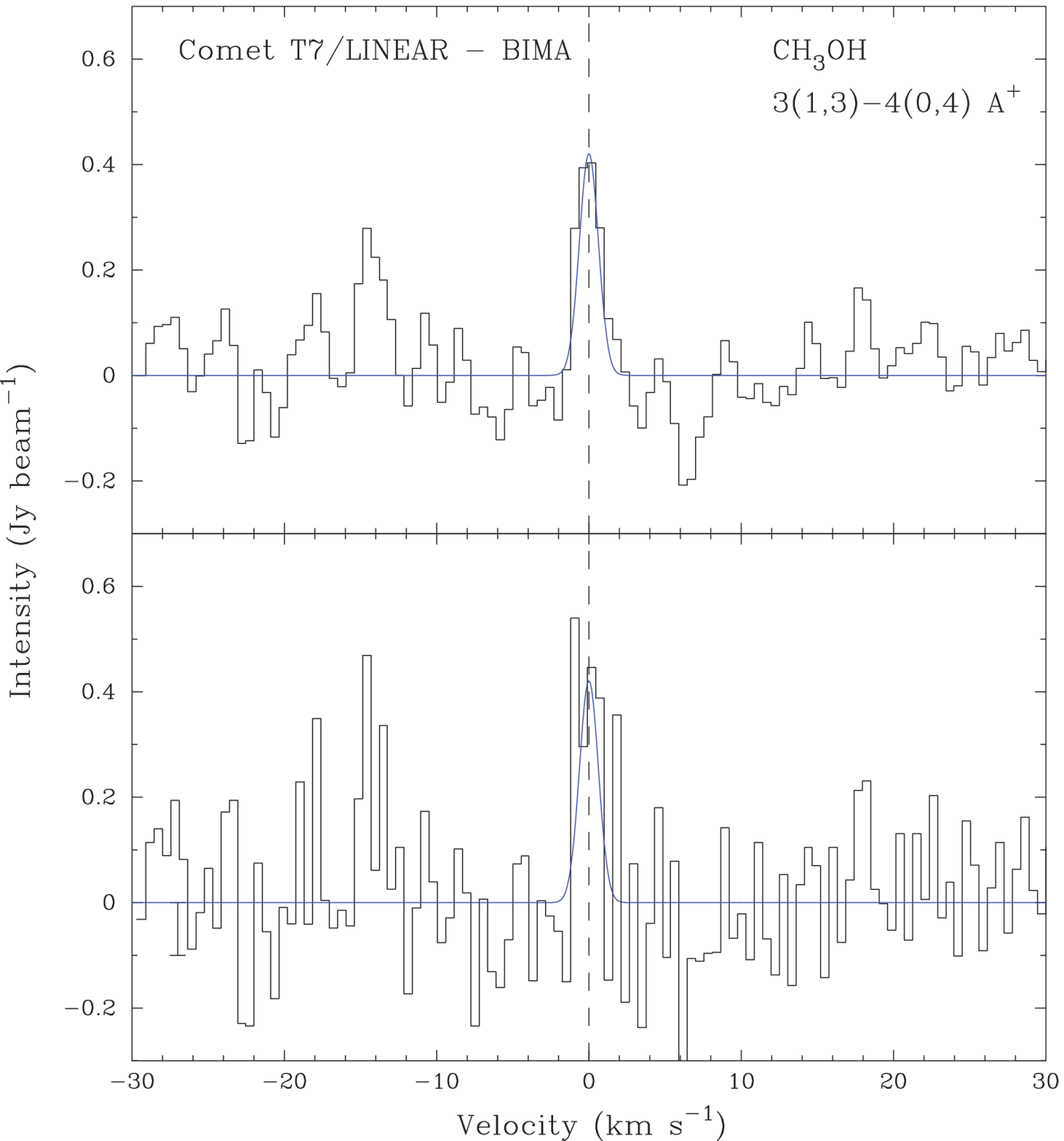}
\centerline{Figure 5.}
\end{figure}
\clearpage
\begin{figure}
\epsscale{1.0}
\plotone{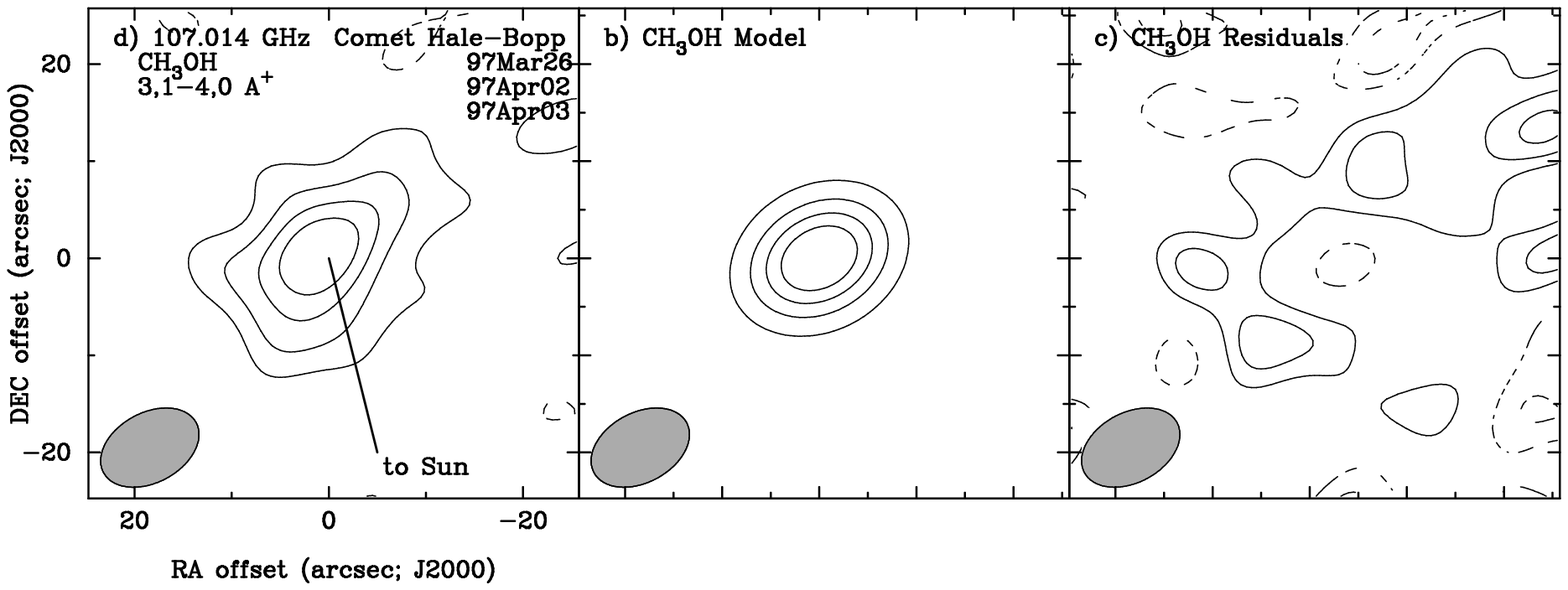}
\centerline{Figure 6.}
\end{figure}

\end{document}